\documentstyle[11pt,aaspp4,epsf]{article}

\newcommand\D{{\cal D}}
\def\psim{\lower.5ex\hbox{$\; \buildrel \propto \over \sim \;$}}
\def\D{{\cal D}}
\def\to{t_{\rm obs}}
\def\eo{\epsilon_{\rm obs}}
\def\muo{\mu_{\rm obs}}

\begin{document}

\title{On Spectral and Temporal Variability in Blazars\\ and Gamma Ray Bursts }

\author{Charles D. Dermer}

\affil{E. O. Hulburt Center for Space Research, Code 7653,\\
       Naval Research Laboratory, Washington, DC 20375-5352}

\begin{abstract}

A simple model for variability in relativistic plasma outflows is studied, in which
nonthermal electrons are continuously and uniformly injected in the comoving frame over a
time interval $\Delta t$.  The evolution of the electron distribution is assumed to be
dominated by synchrotron losses, and the energy- and time-dependence of the synchrotron
and synchrotron self-Compton (SSC) fluxes are calculated for a power-law electron
injection function with index $s = 2$.  The mean time of a flare or pulse measured at
photon energy $E$ with respect to the onset of the injection event varies as $ E^{-1/2}$
and $ E^{-1/4}$ for synchrotron and SSC processes, respectively, until the time
approaches the limiting intrinsic mean time $(1+z)\Delta t/(2\cal D)$, where $z$ is the
redshift and $\cal D$ is the Doppler factor.  This dependence is in accord with recent
analyses of blazar and GRB emissions, and suggests a method to discriminate between
external Compton and SSC models of high-energy gamma radiation from blazars and GRBs. 
The qualititative behavior of the X-ray spectral index/flux relation observed from BL Lac
objects can be explained with this model.  This demonstrates that synchrotron losses are
primarily responsible for the X-ray variability behavior and strengthens a new test for
beaming from correlated hard X-ray/TeV observations.

\end{abstract}

\keywords{BL Lacertae objects: general --- galaxies: jets --- gamma rays: bursts ---
radiation mechanisms: nonthermal}

\section{Introduction}

Space-based observatories such as {\it ASCA}, {\it RXTE}, {\it Beppo-SAX} and the {\it
Compton Gamma Ray Observatory}, and ground-based air Cherenkov telescopes such as the
{\it Whipple Observatory} and {\it HEGRA}, have gathered data from blazars and GRBs of
sufficient quality to allow combined temporal and spectral analyses. 
A general trend is becoming apparent in blazar studies (see
Wagner \markcite{Wagner97}1997 and Shrader \& Wehrle
 \markcite{sw97}1997 for reviews): variability of optical 
and $> 100$ MeV emission in flat spectrum sources
such as 3C 279 (e.g., Hartman et al.
\markcite{Hartman96}1996) and BL Lacertae (Bloom et al.
\markcite{Bloom97}1997), and of X-ray and TeV emission in the BL Lac objects Mrk 421 and
Mrk 501 (e.g., Macomb et al.
\markcite{Macomb95}1995; Buckley et al.
\markcite{Buckley96}1996; Catanese et al. \markcite{Catanese97}1997) appears to be
temporally correlated, implying that the same population of electrons
produces both the optical/X-ray and the 100 MeV - TeV emission. In the specific case of
the 1994 May flare of Mrk 421, {\it ASCA} observations (Takahashi et al.
\markcite{Takahashi96}1996) also show that the time lag of photons in the 0.5-2.0 keV
range relative to 5 keV X-rays varies as $E({\rm keV})^{-1/2} - 5^{-1/2}$.  Moreover,
the Mrk 421 flare data follow a well-defined trajectory in a spectral index/flux display,
and this behavior is also found in several other BL Lac objects (e.g., OJ 287, Idesawa et
al. \markcite{Idesawa97}1997; PKS 2155-304, Sembay et al. \markcite{Sembay93}1993)

Recent analyses of GRB data show that the light curves tend to become narrower at
higher energies (Fishman et al. \markcite{Fishman92}1992), and that the decaying phases of
GRB light curves are generally longer than the rising phases (Link, Epstein, \&
Priedhorsky \markcite{Link93}1993).  Fenimore et al. (\markcite{Fenimore95}1995) showed
that the autocorrelation function of 45 bright BATSE GRBs, aligned at their peak fluxes,
displays an energy dependent width
$\propto E^{-0.4}$. In the specific case of GRB 960720, which displays a single
well-defined pulse, an energy-dependent duration $\propto E^{-0.46}$ is found (Piro et
al. \markcite{Piro98}1998).   

The radio/optical continuum of blazars is thought to be nonthermal synchrotron
radiation, and blast wave models of GRBs also indicate that the hard X-ray/soft gamma-ray
continuum from GRBs is produced by the same process (e.g.,  M\'esz\'aros, Rees, \&
Papathanassiou \markcite{MRP94}1994; Tavani \markcite{Tavani1996}1996; Waxman
\markcite{Waxman97}1997; Vietri \markcite{Vietri97}1997).  The origin of the high energy
gamma-ray emission in blazars is generally thought to originate from Compton processes,
though it remains unclear whether internal synchrotron photons or 
photons produced outside the jet represent the dominant soft photon source.  The
high-energy radiation from GRBs might also originate from Compton-scattering processes
(see, e.g., Hurley et al. \markcite{Hurley94}1994; M\'esz\'aros et al.
\markcite{MLR93}1993).  The likelihood that synchrotron emission from cooling nonthermal
electrons can  explain some of the previously mentioned trends has been discussed in
various approximations (e.g., Tashiro et al. \markcite{Tashiro95}1995; Takahashi et al.
\markcite{Takahashi96}1996; Tavani \markcite{Tavani1996}1996; Kazanas, Titarchuk, \& Hua
\markcite{Kazanas98}1998; Kirk, Rieger, \& Mastichiadis \markcite{krm98}1998), but without
a consideration of the associated SSC emission and a full treatment of the relativistic
boost and energy normalization for the injected nonthermal electrons.

In this {\it Letter}, we analyze the simplest possible model that retains the essential
physics of nonthermal electron injection and cooling in relativistic plasma outflows. 
Even so, it goes a long way towards explaining the observed trends, and in addition
suggests a fruitful avenue of research which could help to discriminate between models
of high-energy blazar emission, to determine the origin of the high-energy radiation in
GRBs, and to strengthen a new beaming test for blazar radiation (Catanese et al.
\markcite{Catanese97}1997). 

\section{Analysis}

Assume that nonthermal electrons with Lorentz factors $\gamma_1 \leq \gamma_i \leq
\gamma_2$ are injected uniformly throughout a spherical emission region with radius $l$,
and that the electron injection spectrum maintains a constant amplitude and power-law
form with injection index $s$ between comoving times $t_1
\leq t \leq t_1+\Delta t$.  Note that causality restricts $\Delta t$ to be no shorter
than $\sim l/c$, though the injection duration could be much longer if the time scale
$\Delta t$ is set by some intrinsic process which energizes the nonthermal electrons on
a longer time scale.   If the total energy injected in nonthermal electrons during the
injection event is
${\cal E}_e$(ergs), then the injection function at time
$t_i$ is given by
$$\dot N_e(\gamma_i, t_i) = {(2-s){\cal E}_e\over (\gamma_2^{2-s}-\gamma_1^{2-s})m_ec^2
\Delta t}\;\gamma_i^{-s}\;\Theta(\gamma_i;\gamma_1,\gamma_2)\;\Theta(t_i;t_1,t_1+\Delta
t)\;,\eqno(1)$$
\noindent where the Heaviside function $\Theta(x;a,b) = 1$ if $a\leq x < b$ and
$\Theta(x;a,b) = 0$ otherwise.  Equation (1) is readily generalized to time-varying
injection events and, given some work, to cases where the injection occurs nonuniformly
throughout the volume of the plasmoid.

The synchrotron energy-loss rate of isotropic relativistic electrons in a randomly
oriented constant magnetic field with mean field strength $H$(Gauss) is given by the
well-known expression
$-\dot\gamma = \nu_0\gamma^2$, where $\nu_0 =  c \sigma_{\rm T} H^2/(6\pi m_e c^2) =
1.29\times 10^{-9} H^2$ s$^{-1}$.  Here we assume that synchrotron losses dominate
adiabatic losses and Compton losses.  The former condition holds for electrons with
$\gamma \gtrsim 8\times 10^3 \beta_{\rm exp}/[H^2({\rm G}) (l/3\times 10^{15}~{\rm
cm})]$, where $\beta_{\rm exp} $ is the expansion speed in units of $c$.  The latter
condition requires that the photon energy density in the comoving frame is less than the
magnetic field energy density.  

When these conditions hold, the synchroton energy-loss
equation is easily solved to give $\gamma(t) = [\gamma_i^{-1}+\nu_0(t-t_i)]^{-1}$  The
evolving electron distribution function at time $t$ is therefore given by
$$N_e(\gamma;t) = \int _0^t dt_i \;\dot N_e (\gamma,t_i) = K_e\gamma^{-2}
\int_{\max(0,t-T)}^{\min(t,\Delta t)}\;dt_i\;[\gamma^{-1}-\nu_0(t-t_i)]^{s-2}\;,\eqno(2)$$
\noindent where $K_e$ is the coefficient preceeding $\gamma_i^{-s}$ in equation (1) and
$T = T(\gamma) \equiv \nu_0^{-1}(\gamma^{-1}-\gamma_2^{-1})$ is the 
time scale for an electron injected with $\gamma_2$ to reach Lorentz factor $\gamma$
through synchrotron losses.  The expression on the rhs of equation (2),
valid for $\gamma_1\leq \gamma\leq \gamma_2$, is obtained by noting that
$|d\gamma_i/d\gamma | = \gamma_i^2/\gamma^2$.  We have also set $t_1= 0$ without loss of
generality.  It is straightforward to generalize equation (2) for particle escape (see
Tashiro et al. 1995). 

We also assume that there is a separation between the
acceleration/injection and cooling processes.  This separation holds when the cooling time
scale is much longer than the acceleration time scale, but can limit the highest
energies to which electrons are accelerated during the main
portion of a flare, as shown by Kirk et al. (\markcite{krm98}1998).  Particle
acceleration can also influence spectral and temporal evolution long after
the main flaring behavior has ended, though its effect is probably negligible if
the flares exhibit sharply defined pulses  which return to their quiescent levels at
high energies.

The spectral flux S(ergs s$^{-1}$ cm$^{-2}$ $\epsilon_{\rm obs}^{-1}$) observed at energy
$E=m_ec^2\epsilon_{\rm obs}$ and observer time $t_{\rm obs}$ from a plasmoid at redshift
$z$ and luminosity distance $d_L$ is given by
$$S(\epsilon_{\rm obs};\Omega_{\rm obs},t_{\rm obs}) = {{\cal D}^3 (1+z)\over 4\pi
d_L^2}\; J(\epsilon,t)\;.\;\eqno(3)$$
\noindent Here $\Omega_{\rm obs} = (\mu_{\rm obs},\phi_{\rm obs})$ specifies the observer
direction, where $\theta_{\rm obs}=\arccos\mu_{\rm obs}$ is the angle between the jet
axis and the direction to the observer, and $J$(ergs s$^{-1}$ $\epsilon^{-1}$) is the
spectral emissivity integrated over the volume of the plasmoid.  Equation (3) is valid
for radiation which is emitted isotropically in the comoving frame, and applies to
synchrotron and SSC processes (see Reynolds \markcite{Reynolds91}1991 and Dermer
\markcite{Dermer95}1995 for cases where this does not hold).  The expressions $\epsilon =
(1+z)\epsilon_{\rm obs}/{\cal D}$ and $dt = {\cal D}dt_{\rm obs}/(1+z)$ relate the
comoving and observed photon energies and the comoving and measured differential time
elements, respectively, where the Doppler factor
$\D\equiv[\Gamma(1-B\muo)]^{-1}$ and $B\equiv (1-1/\Gamma^2)^{1/2}$, and $\Gamma$ is the
Lorentz factor of the plasmoid.  We assume that $\Gamma$ remains constant throughout the
flaring event, which implies that if if swept-up matter enerigizes the plasma, then the
mass of the swept-up matter is $\ll {\cal M}/\Gamma$, where ${\cal M}$ is the total mass
of the particles in the plasmoid. A self-consistent treatment of plasmoid dynamics when
the inertia of external matter swept up by the plasmoid cannot be neglected has been
treated by Chiang \& Dermer (\markcite{cd98}1998).

In the $\delta$-function approximation for synchrotron emission, which is a good
approximation away from the cutoffs of the synchrotron spectrum produced by the endpoints
of the electron distribution, we have 
$$ J_{\rm syn}(\epsilon,t) = {2c\sigma_{\rm T}u_H\over
3\epsilon_H}\;({\epsilon\over\epsilon_H})^{1/2}\;N_e[({\epsilon\over\epsilon_H})^{1/2};t]\;\eqno(4)$$
(e.g., Dermer, Sturner, \& Schlickeiser \markcite{Dermer97}1997). The term $u_H =
H^2/8\pi$ is the magnetic-field  energy density, and $\epsilon_H =
H/4.414\times10^{13}{\rm G}$ is the dimensionless electron plasma frequency.

After substituting equation (4) into equation (3) and using equation (2) for the electron
spectrum, we find for the case $s=2$ that
$$S_{\rm syn}(\epsilon_{\rm obs};\Omega_{\rm obs},t_{\rm obs}) =  {C_{\rm syn}\over
\eo^{1/2}}\;\{ \min(t,\Delta
t)-\max[0,t-T(\sqrt{\epsilon\over\epsilon_H})]\}\;,\;\eqno(5)$$
\noindent where the coefficient $C_{\rm syn} =  \D^{7/2}(1+z)^{1/2} K_e c\sigma_{\rm T}
u_H/(6\pi d_L^2\epsilon_H^{1/2})$. The case $s=2$ is arguably the most interesting case
since, according to simple shock acceleration theory, first-order Fermi acceleration by a
strong shock produces a particle spectrum with $s\cong 2$ in nonrelativistic monatomic
gases, provided that nonlinear feedback of the accelerated particles can be neglected. 
More general injection functions can be considered in more detailed treatments.

The volume-integrated SSC emissivity in the $\delta$-function approximation is given by
$$J_{\rm SSC}(\epsilon,t) = {c\sigma_{\rm T}^2 u_{\rm H} \epsilon^{1/2}\over 3\pi
\epsilon_H^{3/2} l^2}\;\int_{\epsilon_H}^{\min(\epsilon,\epsilon^{-1})}
d\epsilon'\;\epsilon'^{-1}\;N_e(\sqrt{\epsilon'\over
\epsilon_H};t)\;N_e(\sqrt{\epsilon\over
\epsilon'};t)\;\eqno(6)$$ 
(Dermer et al. \markcite{Dermer97}1997).  The upper limit
on the integral restricts the scattering to the Thomson regime.  After substituting
equation (6) into equation (3) and making use of equation (2) for the electron
distribution, one obtains, again for the case $s=2$, the result
$$S_{\rm SSC}(\eo;\Omega_{\rm obs},\to) = {C_{\rm SSC}\over
\eo^{1/2}}\int_{\epsilon_H}^{\min(\epsilon,\epsilon^{-1})}d\epsilon'\;\epsilon'^{-1}\;
\{\min(t,\Delta t) -\max[0,t-T(\sqrt {\epsilon'\over\epsilon_H})]\}\;$$
$$~~~~~\;\times\{\min(t,\Delta t) -\max[0,t-T(\sqrt {\epsilon\over\epsilon'})
]\}\;.\eqno(7)$$ The coefficient $C_{\rm SSC} = C_{\rm syn} K_e \sigma_{\rm T}/(2\pi
l^2 )$. 
 
\section{Results and Discussion} 

Figure 1 shows the time-dependence of the synchrotron (Fig. 1a) and SSC (Fig. 1b)
spectral fluxes for a variety of different observing energies, obtained by numerically
solving equations (5) and (7), respectively.  Because $S\propto \eo^{-1/2}$
from uncooled electrons injected with $s=2$, the spectral flux is multiplied by
$\eo^{1/2}$ for clarity of presentation on a linear scale. In this calculation, we let
$H = 0.1$ Gauss, $\gamma_1 = 10$, $\gamma_2 = 10^8$, $z=0.1$, $\D = 10$, and $\Delta t =
10^5$ s, as might be appropriate for a blazar flare.

We wish to determine the mean time $\langle t(\eo )\rangle$ of the
radiation observed at different energies, measured with respect to the onset of the
injection event.  This is given by the expression
$$\langle t(\eo )\rangle =  {\int _{-\infty}^{\infty} d\to \cdot \to \cdot 
S(\eo;\Omega_{\rm obs},\to )\over \int _{-\infty}^{\infty} d\to\cdot S(\eo;\Omega_{\rm
obs},\to ) }  \,.\;\eqno(8)$$
\noindent Note that as defined here, a Heaviside (or boxcar) function has a mean
time equal to one-half its temporal duration.  Equation (8) can be easily generalized
for for higher moments of the temporal profile giving, for example, the FWHM duration of a
flare.

After substituting equation (5) into equation (8), one obtains two cases depending on
whether $T\leq \Delta t$ of $T > \Delta t$.  When $\gamma_2\rightarrow \infty$, the same
result is found in both cases, namely
$$\langle t (\eo )\rangle _{\rm syn} = {1+z\over 2\D} \;\big[ \Delta t
+\nu_0^{-1}\sqrt{{\D\epsilon_H\over (1+z)\eo}} \;\big] 
\,.\;\eqno(9)$$

Equation (9) provides a convenient expression for fitting energy-dependent time-lag data,
such as the Mrk 421 flare (Takahashi et al. \markcite{Takahashi96}1996), or the
energy-dependent GRB widths measured by Fenimore et al. (\markcite{Fenimore95}1995) and
Piro et al. (\markcite{Piro98}1998), though the appropriate moment analysis should be
used in more detailed treatments. Equation (9) indicates that the energy-dependence of the
mean time varies more slowly than
$E^{-0.5}$ when the light-crossing time $l/c$ or duration of the energization event are
comparable to or longer than the comoving electron cooling time scale.  Fitting high
quality data to equation (9) to determine the photon energy
$\bar\eo$ where the two branches of the expression intersect, and using the shortest
variability time scale  $\delta\to^{\rm min} = (1+z)\Delta t/\D $ observed at $\eo \gg
\bar\eo$, we derive the plasmoid magnetic field
$$H({\rm G}) \cong 0.8\;\{ {(1+z)\over \D \bar E({\rm keV}) [\delta \to^{\rm min}({\rm
hr})]^2 }\}^{1/3}\;.\eqno(10)$$ 
Causality implies that $l\lesssim c\D \delta \to^{\rm
min}/(1+z)$. It will be important in future studies to understand how equations (9) and
(10) are modified for a non-spherical plasmoid geometry and (especially for GRBs; see,
e.g., Fenimore, Madras, \& Nayakshin \markcite{Fenimore96}1996) a blast wave geometry.

Expressions similar to equations (9) and (10) have been noted (e.g., Tashiro et al.
\markcite{Tashiro95}1995; Takahashi et al. \markcite{Takahashi96}1996; Tavani
\markcite{Tavani96}1996; Buckley et al. \markcite{Buckley98}1998; Catanese et al.
\markcite{Catanese97}1997; Kazanas et al.
\markcite{Kazanas98}1998), but this treatment provides a precise fitting function for
analysis of data when $s= 2$ and yields a generalization for arbitrary values of
$s$.  In
the paper by Catanese et al. (\markcite{Catanese97}1997), moreover, it is noted that
correlated X-ray/TeV data imply an upper limit on $H$ because the electrons producing the
highest energy synchrotron emission have Lorentz factors
$\gtrsim 2\times 10^6 E_{\rm C}({\rm TeV}) (1+z)/\D$, where $E_{\rm C}({\rm TeV})$ is the
measured energy in TeV of the highest-energy gamma-rays. Synchrotron emission 
correlated with the TeV flux requires that the electrons radiate in a magnetic field
at least as great as $H({\rm G}) \cong 11 \epsilon_{\rm obs, syn}\D/[ E^2_{\rm C}({\rm
TeV}) (1+z)]$, where $\epsilon_{\rm obs, syn}$ is the measured dimensionless energy of the
highest energy synchrotron photons produced by the electrons which produce the
TeV radiation. When compared with the value of
$H$ inferred through equations (9) and (10), we obtain an
expression for the Doppler factor, given by 

$$\D \cong 1.7\;{(1+z)  [E_{\rm C}({\rm TeV})]^{3/2}\over \bar \epsilon_{\rm
obs}^{1/4}({\rm keV}) (\delta \to^{\rm min})^{1/2}(\epsilon_{\rm obs,
syn})^{3/4}}\;.\eqno(11)$$ A lower limit to $\D$ is obtained if the TeV flux does not
exhibit a clear cutoff due to the high-energy cutoff in the the electron distribution
function.

The inset to Fig. 1a shows the energy dependence of the mean time $\langle t (E)\rangle $
for the synchrotron and SSC processes.  As can be seen by examining Fig. 1b, the SSC
emission decays more slowly than the synchrotron process, yielding an energy dependence
$\propto E^{-1/4}$ until the shortest mean time $(1+z)\Delta t/2\D $ is reached,
which occurs at much higher energies than for the synchrotron process.  Thomson
scattering of external photons, having an energy loss rate of the same form as the
synchrotron energy loss rate, also gives $\langle t(\eo )\rangle \propto
E^{-1/2}$ for $s=2$.  When more sensitive blazar $\gamma$-ray observations become
available with the upcoming
$\it INTEGRAL$ and $\it GLAST$ missions, the energy-dependence of the mean time
or duration of blazar flares can be used to determine whether SSC or ECS processes
dominate in specific sources or between different source classes. This test might also be
possible for bright TeV flares from BL Lac objects given the steadily improving
sensitivity of air Cherenkov telescopes. More detailed studies need
to be performed which take into account Klein-Nishina effects and general values of $s$.

Figure 2 shows the evolution of the synchrotron spectrum at different observing times,
using the same parameters as in Figure 1.  At a fixed photon energy, the flux rises due
to the accumulation of injected nonthermal electrons until cooling plays an important role
in depleting the electron spectrum (cf. Dermer \& Chiang \markcite{Dermer98}1998).  The
effects of cooling are seen earliest at the highest photon energies, and cause a
softening of the spectrum.  Consequently, the spectral flux at a given photon energy
displays a hard spectrum while its intensity is increasing, since the electrons that are
radiating this emission have not yet felt the effects of cooling.  When synchrotron
cooling becomes important, the flux begins to fall rapidly.  This behavior is shown by
the solid curve (1) in the inset to Figure 2, which displays the clockwise evolution of
the 2-10 keV spectral index as a function of a quantity proportional to the 2 keV spectral
flux.  In reality, there will be some level of background which will be reached.  This is
crudely modeled here by adding an underlying spectral component with spectral flux
$S_{\rm bg}(\eo )=K_{\rm bg}\eo^{-0.5}$, where $K_{\rm bg} = 1$  and
$5\times 10^4$ ergs cm$^{-2}$ for the dashed curve (2) and dotted curve (3), respectively.

The behavior illustrated in the inset to Fig. 2 is in qualitative agreement with 
evolutionary tracks of the spectral index/flux observed from some blazars, as noted
in the Introduction (see also Kirk et al.\ \markcite{krm98}1998). Precise fitting of such
tracks will require extending this model for general injection indices and for various
forms for the background emission. The well-known hard-to-soft evolution of GRB pulses
(e.g., Norris et al.\ \markcite{Norris86}1986) could be a manifestation of this effect,
and the approach outlined here can also be used to analyze the fluence dependence of the
peak of the GRB $\nu F_\nu$ spectrum (Liang \& Kargatis \markcite{Liang96}1996).

In summary, a very simple model has been presented for the observed nonthermal
synchrotron and SSC emission emitted by cooling nonthermal electrons which are injected
over a comoving time interval $\Delta t$ into plasma with relativistic bulk motion.  In
spite of the model's simplicity, empirical trends that have become better defined through
recent combined temporal and spectral analyses of data from blazars and GRBs
are qualitatively understood.  Straightforward generalizations, necessary for detailed
fits to data, were indicated thoughout the analysis and will be treated in future
studies. 

\acknowledgements I thank Ed Fenimore and Michael Catanese for conversations and
questions that focused my attention on this problem. Comments by J.
Chiang and the referee are acknowledged. This research was supported by the Office of
Naval Research and the {\it Compton Gamma Ray Observatory} Guest Investigator Program.

\eject

\begin{figure}
\centerline{\epsfxsize=12.5cm \epsfbox{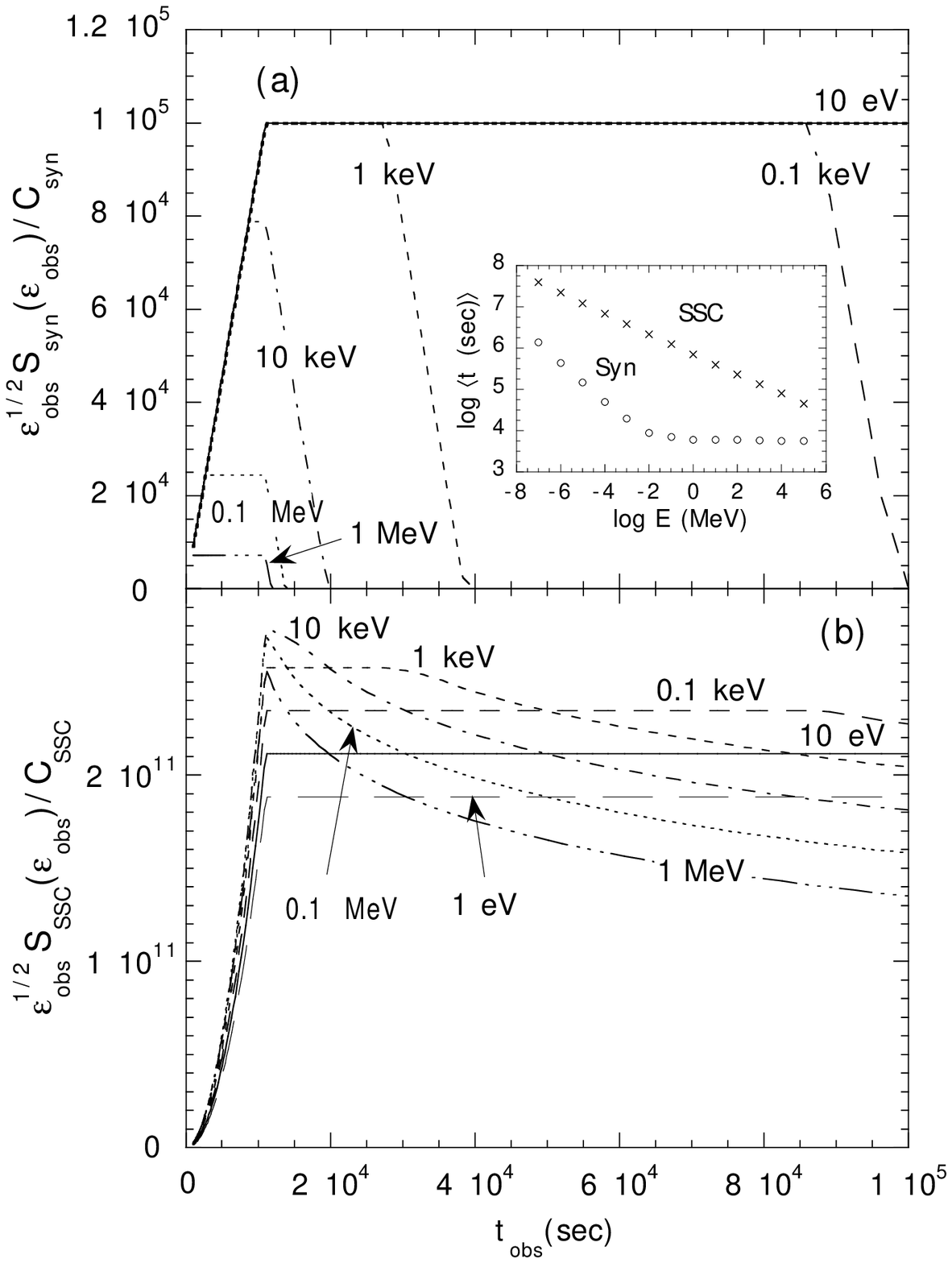}}
\caption{Synchrotron (Fig. 1a) and SSC (Fig. 1b) fluxes as a function of time at
different photon energies.  Nonthermal electrons with injection index $s =2$ are injected
for $\Delta t =10^5$ seconds into a spherical plasmoid at
redshift $z = 0.1$, and with Doppler factor $\D = 10$ with respect to the observer.  Other
parameters are given in the text. The mean time of the fluxes as a function of observer
energy, defined by equation (8), are shown in the inset.}

\end{figure}
                 
\eject

\begin{figure}
\centerline{\epsfxsize=12.5cm \epsfbox{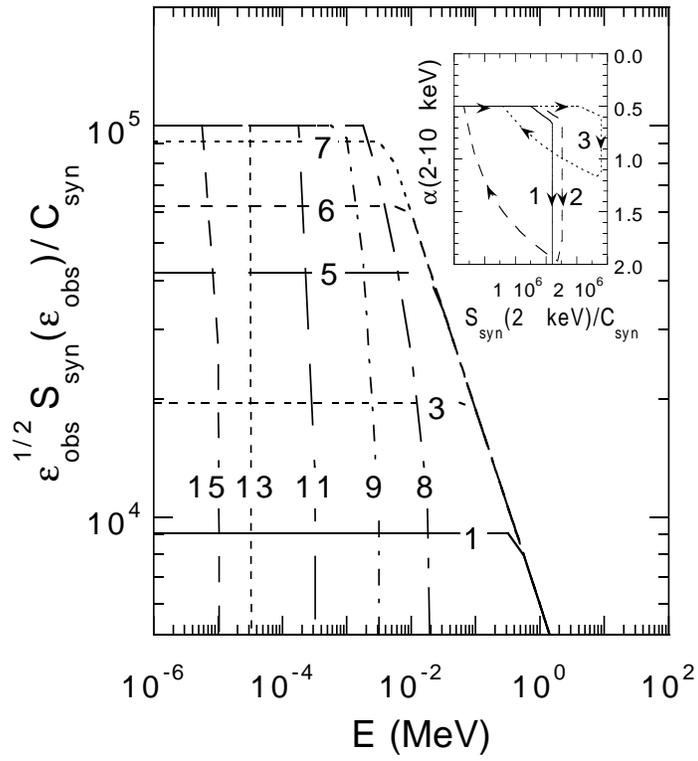}}
\caption{Synchrotron fluxes as a function of observing energy  at different observing
times $\to ({\rm s}) = [(1+z)\Delta t/\D) ]\times 10^{(j-7)/6}$, using the same parameters
as in Figure 1. The curves are labeled by $j$. The inset shows the tracks followed by the
2-10 keV spectral index versus a quantity proportional to the 2 keV spectral flux. 
Various levels of background fluxes, as described in the text, are included in the
calculations of the dashed and dotted curves.   }

\end{figure}
                 
\end{document}